\documentclass[preprint,showpacs,preprintnumbers,amsmath,amssymb, prb]{revtex4}

\usepackage{graphicx}
\usepackage{dcolumn}

\usepackage[latin1]{inputenc}
\usepackage{amssymb, bm}
\usepackage{setspace}
\usepackage{lineno}

\def\0{\phantom{0}}
\def\.{\phantom{.}}

\begin{document}

\preprint{APS/123-QED}

\title{Comment on ``An optimized potential for carbon dioxide'' \\{} [J. Chem. Phys. 122, 214507] (2005)}

\author{Thorsten Merker}
\affiliation{Laboratory of Engineering Thermodynamics, University of Kaiserslautern, 67633 Kaiserslautern, Germany}

\author{Jadran Vrabec \footnote{ Author to whom correspondence should be adressed: Tel.: +49-711/685-66107, Fax:
+49-711/685-66140, Email: vrabec@itt.uni-stuttgart.de}}
\affiliation{Institute of Thermodynamics and Thermal Process Engineering, University of Stuttgart, 70550 Stuttgart, Germany}

\author{Hans Hasse}
\affiliation{Laboratory of Engineering Thermodynamics, University of Kaiserslautern, 67633 Kaiserslautern, Germany }

\date{\today}	

\begin{abstract}
A molecular model for carbon dioxide is assessed regarding vapor-liquid equilibrium properties. Large deviations, being above 15~\%, are found for vapor pressure and saturated vapor density in the entire temperature range.
\end{abstract}

\pacs{Valid PACS appear here}


\maketitle

In a recent article, Zhang and Duan \cite{zhang2005} presented a new potential model for carbon dioxide ($\mathrm{CO_2}$). It consists of three Lennard-Jones (LJ) sites to account for repulsion and dispersion and three distributed partial charges to describe the quadrupolar interaction. The molecular model is rigid and rotationally symmetric around the molecular axis. In that work \cite{zhang2005}, simulation results on vapor-liquid equilibria (VLE), radial distribution function and self-diffusion coefficient have been reported for the new $\mathrm{CO_2}$ model which are in excellent agreement with experimental data. They also compared to results from different other models from the literature. The new model is found "to be superior to the previous models in general".

For the VLE properties, deviations between model and experimental data are reported to be \cite{zhang2005}:  0.7~\% for vapor pressure,  0.1~\% for saturated liquid density, 2.3~\% for saturated vapor density, and 1.9~\% for heat of vaporization. 

Particularly for vapor pressure, which is the most sensitive of those properties, the authors claim to achieve an average accuracy of better than 1~\% over the entire temperature range with a molecular model which is noteworthy for any molecular model.

As there is widespread scientific interest in $\mathrm{CO_2}$, the model by Zhang and Duan \cite{zhang2005} was employed for subsequent work \cite{zhang2007,Eslami2007,idrissi2007,idrissi2006,zhang2006}. It should be pointed out that no VLE data were published in \cite{zhang2007,Eslami2007,idrissi2007,idrissi2006,zhang2006}. We recently tested the $\mathrm{CO_2}$ model from Zhang and Duan \cite{zhang2005} and found results that strongly deviate from these reported by Zhang and Duan \cite{zhang2005}. The new simulation results also strongly deviate from experimental data, particularly the vapor pressure and the saturated vapor density.

The present assessment was made on the basis of two simulation tools that employ different methods to determine VLE. Firstly, the Grand Equilibrium method \cite{jadran} was used as implemented in our simulation tool $ms2$ \cite{deublein} and, secondly, the Gibbs ensemble \cite{Gibbs} was used as implemented in the freely available simulation tool TOWHEE \cite{TOWHEE}. Both programs have proven to be correct e.g. in the recent Industrial Fluid Property Simulation Challenge for the case of ethylene oxide which is very similar to the present case \cite{eckl}.

For our approach with $ms2$, molecular dynamics (MD) simulations were performed in the liquid phase containing 1024 molecules. After a sufficient equilibration period, the chemical potential was calculated by Widoms's insertion method \cite{widom} over 300~000 time steps. According to the Grand Equilibrium method \cite{jadran}, the dew point was sampled with Monte-Carlo (MC) simulations where approximately 500 molecules were used. The simulation details were similar to those published in \cite{eckl} and are not repeated here.

With TOWHEE \cite{TOWHEE}, Gibbs ensemble MC simulations \cite{Gibbs} were performed. There, smaller systems were studied, containing 500 molecules. After an equilibration over 5~000 loops without volume and molecules transfer moves, followed by 10~000 loops with these moves, 50~000 production loops were performed. Other simulation details were similar to those published in \cite{ozgur} and are not repeated here.

\begin{figure}[ht]
\includegraphics[scale=0.7]{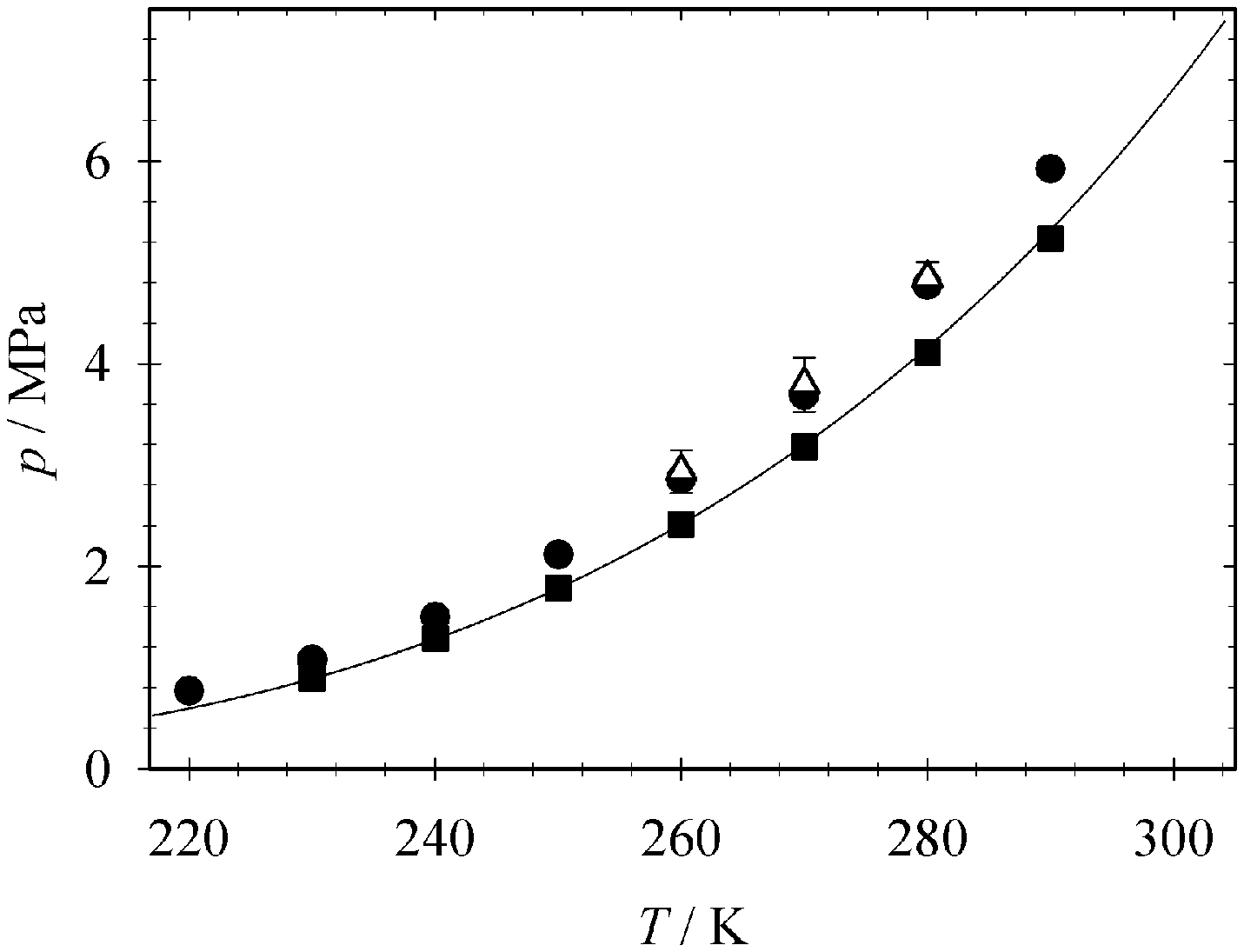}
\bigskip
\caption[Vapor pressure of carbon dioxide. {\large $\bullet$}~present results Grand Equilibrium method \cite{jadran}, {\small $\triangle$}~present results Gibbs ensemble method \cite{Gibbs}, {\small $\blacksquare$}~results by Zhang and Duan \cite{zhang2005}, {---}~reference EOS \cite{data}.]{ \label{fig_vle_p}}
\end{figure}

\begin{figure}[ht]
\includegraphics[scale=0.7]{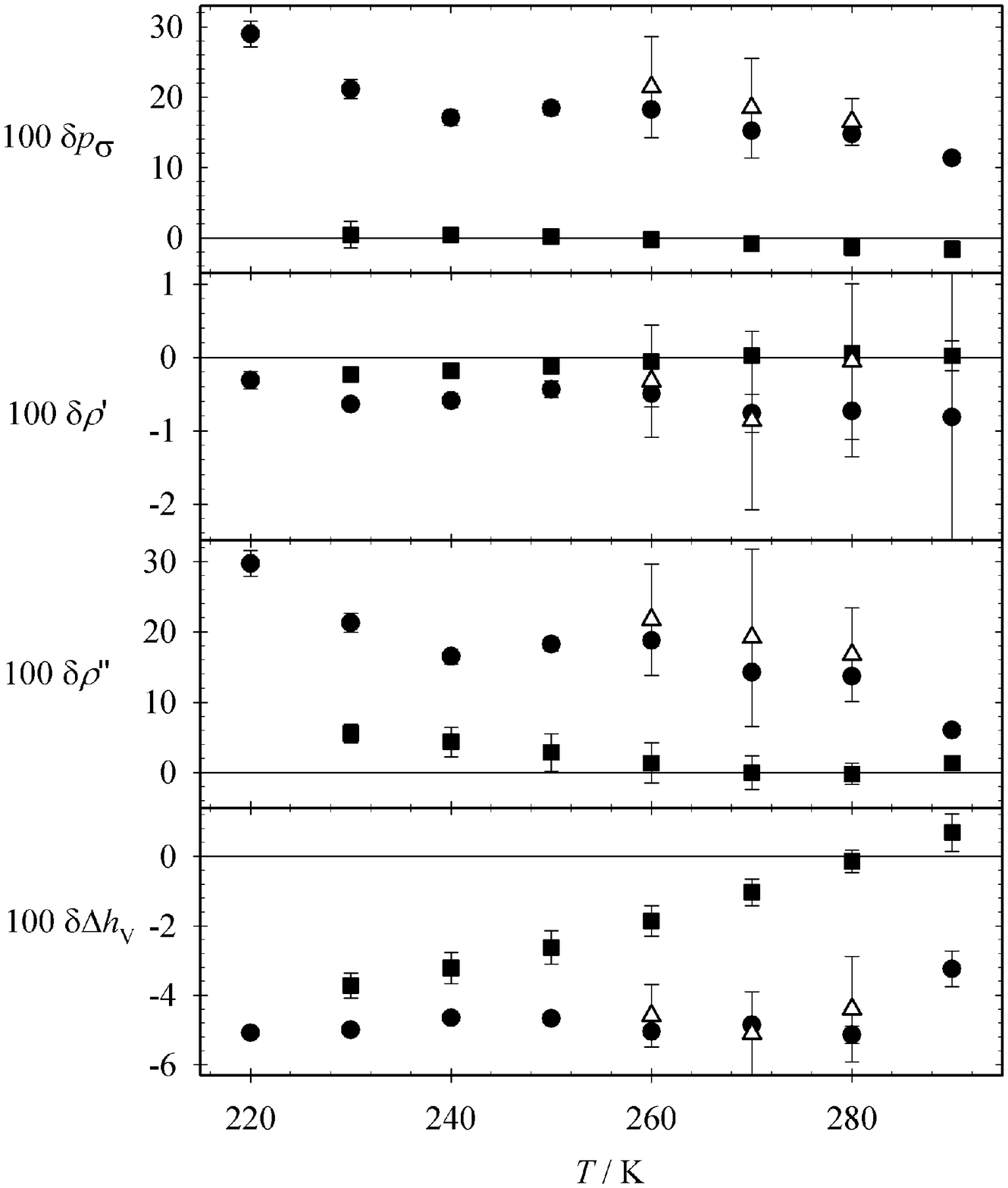}
\bigskip
\caption[Relative deviations of vapor-liquid equilibrium properties between simulation and reference EOS \cite{data} ($\delta z = (z_{\mathrm{sim}} - z_{\mathrm{eos}}) / z_{\mathrm{eos}}$). {\large $\bullet$}~present results Grand Equilibrium method \cite{jadran}, {\small $\triangle$}~present results Gibbs ensemble method \cite{Gibbs}, {\small $\blacksquare$}~results by Zhang and Duan \cite{zhang2005}, {---}~reference EOS \cite{data}.]{\label{fig_vle_abw}}
\end{figure}

The Present simulation results are compared to those reported by Zhang and Duan \cite{zhang2005} in Table \ref{tab_vle1} as well as in Figures \ref{fig_vle_p} and \ref{fig_vle_abw}, which also contains results from a highly accurate reference data for $\mathrm{CO_2}$ \cite{data} recommended by the National Institute of Science and Technology \cite{NIST}. Figure 2 shows deviation plots where it can be seen that the present data sets based on the two different simulation methods agree with each other within their (combined) error bars throughout. However, they are significantly off the data by Zhang and Duan \cite{zhang2005} which coincide excellently with the experiment.

The present simulation data (Grand Equilibrium) show the following average deviations from experimental data: 18~\% for vapor pressure, 0.6~\% for saturated liquid density, 17~\% for saturated vapor density and 4.6~\% for heat of vaporization. Only the saturated liquid density by Zhang and Duan \cite{zhang2005} and the heat of vaporization are in good agreement with the experimental data, the vapor pressure and saturated vapor density are significantly too high throughout most of the temperature range.

It has to be concluded that the $\mathrm{CO_2}$ model by Zhang and Duan \cite{zhang2005} is not generally superior to previous models.

\bibliography{paper_reviewed}

\begin{thebibliography}{15}
\expandafter\ifx\csname natexlab\endcsname\relax\def\natexlab#1{#1}\fi
\expandafter\ifx\csname bibnamefont\endcsname\relax
  \def\bibnamefont#1{#1}\fi
\expandafter\ifx\csname bibfnamefont\endcsname\relax
  \def\bibfnamefont#1{#1}\fi
\expandafter\ifx\csname citenamefont\endcsname\relax
  \def\citenamefont#1{#1}\fi
\expandafter\ifx\csname url\endcsname\relax
  \def\url#1{\texttt{#1}}\fi
\expandafter\ifx\csname urlprefix\endcsname\relax\def\urlprefix{URL }\fi
\providecommand{\bibinfo}[2]{#2}
\providecommand{\eprint}[2][]{\url{#2}}

\bibitem[{\citenamefont{Zhang and Duan}(2005)}]{zhang2005}
\bibinfo{author}{\bibfnamefont{Z.}~\bibnamefont{Zhang}} \bibnamefont{and}
  \bibinfo{author}{\bibfnamefont{Z.}~\bibnamefont{Duan}}, \bibinfo{journal}{J.\
  Chem.\ Phys} \textbf{\bibinfo{volume}{122}}, \bibinfo{pages}{214507}
  (\bibinfo{year}{2005}).

\bibitem[{\citenamefont{Zhang and Duan}(2007)}]{zhang2007}
\bibinfo{author}{\bibfnamefont{Z.}~\bibnamefont{Zhang}} \bibnamefont{and}
  \bibinfo{author}{\bibfnamefont{Z.}~\bibnamefont{Duan}},
  \bibinfo{journal}{Geochemica et Cosmochima Acta}
  \textbf{\bibinfo{volume}{71}}, \bibinfo{pages}{1164} (\bibinfo{year}{2007}).

\bibitem[{\citenamefont{Eslami and Müller-Plathe}(2005)}]{Eslami2007}
\bibinfo{author}{\bibfnamefont{H.}~\bibnamefont{Eslami}} \bibnamefont{and}
  \bibinfo{author}{\bibfnamefont{F.}~\bibnamefont{Müller-Plathe}},
  \bibinfo{journal}{Macromolecules} \textbf{\bibinfo{volume}{40}},
  \bibinfo{pages}{6431} (\bibinfo{year}{2005}).

\bibitem[{\citenamefont{Idrissi et~al.}(2007)\citenamefont{Idrissi, Damay, and
  Kiselev}}]{idrissi2007}
\bibinfo{author}{\bibfnamefont{A.}~\bibnamefont{Idrissi}},
  \bibinfo{author}{\bibfnamefont{P.}~\bibnamefont{Damay}}, \bibnamefont{and}
  \bibinfo{author}{\bibfnamefont{M.}~\bibnamefont{Kiselev}},
  \bibinfo{journal}{Chem. Phys.} \textbf{\bibinfo{volume}{332}},
  \bibinfo{pages}{139} (\bibinfo{year}{2007}).

\bibitem[{\citenamefont{Idrissi et~al.}(2006)\citenamefont{Idrissi, Longelin,
  and Damay}}]{idrissi2006}
\bibinfo{author}{\bibfnamefont{A.}~\bibnamefont{Idrissi}},
  \bibinfo{author}{\bibfnamefont{S.}~\bibnamefont{Longelin}}, \bibnamefont{and}
  \bibinfo{author}{\bibfnamefont{P.}~\bibnamefont{Damay}}, \bibinfo{journal}{J.
  Chem. Phys.} \textbf{\bibinfo{volume}{125}}, \bibinfo{pages}{224501}
  (\bibinfo{year}{2006}).

\bibitem[{\citenamefont{Zhang and Duan}(2006)}]{zhang2006}
\bibinfo{author}{\bibfnamefont{Z.}~\bibnamefont{Zhang}} \bibnamefont{and}
  \bibinfo{author}{\bibfnamefont{Z.}~\bibnamefont{Duan}},
  \bibinfo{journal}{Geochemica et Cosmochima Acta}
  \textbf{\bibinfo{volume}{70}}, \bibinfo{pages}{2311} (\bibinfo{year}{2006}).

\bibitem[{\citenamefont{Vrabec and Hasse}(2002)}]{jadran}
\bibinfo{author}{\bibfnamefont{J.}~\bibnamefont{Vrabec}} \bibnamefont{and}
  \bibinfo{author}{\bibfnamefont{H.}~\bibnamefont{Hasse}},
  \bibinfo{journal}{Mol. Phys.} \textbf{\bibinfo{volume}{100}},
  \bibinfo{pages}{3375} (\bibinfo{year}{2002}).

\bibitem[{\citenamefont{Deublein et~al.}(2008)\citenamefont{Deublein, Eckl,
  Stoll, Lishchuk, Bernreuther, Vrabec, and Hasse}}]{deublein}
\bibinfo{author}{\bibfnamefont{S.}~\bibnamefont{Deublein}},
  \bibinfo{author}{\bibfnamefont{B.}~\bibnamefont{Eckl}},
  \bibinfo{author}{\bibfnamefont{J.}~\bibnamefont{Stoll}},
  \bibinfo{author}{\bibfnamefont{S.~V.} \bibnamefont{Lishchuk}},
  \bibinfo{author}{\bibfnamefont{M.}~\bibnamefont{Bernreuther}},
  \bibinfo{author}{\bibfnamefont{J.}~\bibnamefont{Vrabec}}, \bibnamefont{and}
  \bibinfo{author}{\bibfnamefont{H.}~\bibnamefont{Hasse}}
  (\bibinfo{year}{2008}), \bibinfo{note}{in preparation}.

\bibitem[{\citenamefont{Panagiotopoulos}(1987)}]{Gibbs}
\bibinfo{author}{\bibfnamefont{A.~Z.} \bibnamefont{Panagiotopoulos}},
  \bibinfo{journal}{Mol. Phys.} \textbf{\bibinfo{volume}{61}},
  \bibinfo{pages}{813} (\bibinfo{year}{1987}).

\bibitem[{TOW()}]{TOWHEE}
\emph{\bibinfo{title}{Towhee monte carlo molecular simulation code}},
  \eprint{http://towhee.sourceforge.net}.

\bibitem[{\citenamefont{Eckl et~al.}(2008)\citenamefont{Eckl, Vrabec, and
  Hasse}}]{eckl}
\bibinfo{author}{\bibfnamefont{B.}~\bibnamefont{Eckl}},
  \bibinfo{author}{\bibfnamefont{J.}~\bibnamefont{Vrabec}}, \bibnamefont{and}
  \bibinfo{author}{\bibfnamefont{H.}~\bibnamefont{Hasse}},
  \bibinfo{journal}{Fluid Phase Equilib.}  (\bibinfo{year}{2008}),
  \bibinfo{note}{doi:10.1016/j.fluid.2008.02.002}.

\bibitem[{\citenamefont{Widom}(1963)}]{widom}
\bibinfo{author}{\bibfnamefont{B.}~\bibnamefont{Widom}}, \bibinfo{journal}{J.
  Chem. Phys.} \textbf{\bibinfo{volume}{39}}, \bibinfo{pages}{2808}
  (\bibinfo{year}{1963}).

\bibitem[{\citenamefont{Özgür Yazaydin and Martin}(2007)}]{ozgur}
\bibinfo{author}{\bibfnamefont{A.}~\bibnamefont{Özgür Yazaydin}}
  \bibnamefont{and} \bibinfo{author}{\bibfnamefont{M.~G.}
  \bibnamefont{Martin}}, \bibinfo{journal}{Fluid Phase Equilib.}
  \textbf{\bibinfo{volume}{260}}, \bibinfo{pages}{195} (\bibinfo{year}{2007}).

\bibitem[{\citenamefont{Span and Wagner}(1996)}]{data}
\bibinfo{author}{\bibfnamefont{R.}~\bibnamefont{Span}} \bibnamefont{and}
  \bibinfo{author}{\bibfnamefont{W.}~\bibnamefont{Wagner}},
  \bibinfo{journal}{J. Phys. Chem. Ref. Data} \textbf{\bibinfo{volume}{25}},
  \bibinfo{pages}{1590} (\bibinfo{year}{1996}).

\bibitem[{NIS()}]{NIST}
\emph{\bibinfo{title}{National institute of standards and technology}},
  \eprint{http://www.nist.gov}.

\end{thebibliography}

\newpage
\begin{turnpage}
\begin{table}[ht]
\noindent
\caption{Vapor-liquid equilibria of carbon dioxide: present simulation results with the Grand Equilibrium method (GE) \cite{jadran} and the Gibbs ensemble (Gibbs) \cite{Gibbs} compared to simulation results by Zhang and Duan (Zhang) \cite{zhang2005} and the reference EOS (eos) \cite{data}. The number in parentheses indicates the statistical uncertainty in the last digit.}
\label{tab_vle1}
\footnotesize{
\begin{center}
\begin{tabular}{c|cccc|cccc|cccc|cccc} \hline\hline
$T$ & $p_{\mathrm{GE}}$ & $p_{\mathrm{Gibbs}}$ & $p_{\mathrm{Zhang}}$ & $p_{\mathrm{eos}}$ & $\rho'_{\mathrm{GE}}$ & $\rho'_{\mathrm{Gibbs}}$ & $\rho'_{\mathrm{Zhang}}$ & $\rho'_{\mathrm{eos}}$ & $\rho''_{\mathrm{GE}}$ & $\rho''_{\mathrm{Gibbs}}$ & $\rho''_{\mathrm{Zhang}}$ & $\rho''_{\mathrm{eos}}$ & $\Delta h^{\mathrm{v}}_{\mathrm{GE}}$ & $\Delta h^{\mathrm{v}}_{\mathrm{Gibbs}}$ & $\Delta h^{\mathrm{v}}_{\mathrm{Zhang}}$ & $\Delta h^{\mathrm{v}}_{\mathrm{eos}}$  \\
K & MPa & MPa & MPa & MPA &  mol/l & mol/l & mol/l & mol/l &  mol/l & mol/l & mol/l & mol/l &  kJ/mol & kJ/mol & kJ/mol & kJ/mol\\ \hline
220 & 0.77(1)  & ---    & ---        & 0.599 & 26.42(3)  & ---    & ---         & 26.497 & 0.466(1) & ---    & ---      & 0.359 & 14.41\0(1)  & ---    & ---         & 15.179\\
230 & 1.08(1)    & ---    & 0.90\0(2)  & 0.893 & 25.48(2)  & ---    & 25.59\0(1)  & 25.646 & 0.641(2)   & ---    & 0.558(7) & 0.529 & 13.717(6)       & ---    & 13.90(5)    & 14.437\\
240 & 1.50(2)  & ---    & 1.29\0(1)  & 1.283 & 24.60(2)  & ---    & 24.697(6)   & 24.742 & 0.882(2) & ---    & 0.79\0(2)& 0.757 & 12.994(7)  & ---    & 13.19(6)    & 13.628\\
250 & 2.11(2)  & ---    & 1.788(2)   & 1.785  & 23.66(3)  & ---    & 23.740(8)   & 23.767 & 1.253(4) & ---    & 1.09\0(3)& 1.06\0 & 12.138(8)  & ---    & 12.40(6)    & 12.733\\
260 & 2.86(2)  & 2.9(2) & 2.41\0(2)  & 2.419 & 22.58(4)  & 22.6(2)& 22.68\0(1)  & 22.697 & 1.738(6)& 1.8(1) & 1.48\0(4)& 1.464 & 11.14\0(1)  & 11.2(1)& 11.51(5)    & 11.728\\
270 & 3.69(2)  &3.8(3)    & 3.18\0(3)  & 3.203 & 21.33(6)  & 21.3(3)    & 21.50\0(1)  & 21.491 & 2.29\0(1) & 2.4(3)    & 2.01\0(5)& 2.008 & 10.06\0(1) & 10.0(2)    & 10.46(4)    & 10.569\\
280 & 4.78(3)  & 4.8(2) & 4.11\0(5)  & 4.161 & 19.9\0(1) & 20.1(2)& 20.09\0(2)  & 20.077 & 3.14\0(2)& 3.2(2) & 2.76\0(4)& 2.766 &\08.71\0(2) &\08.8(1)& \09.17(3)   & \09.183\\
290 & 5.92(4)  & ---    & 5.23\0(6)  & 5.318  & 18.1\0(4)   & ---    & 18.29\0(4)  & 18.284 & 4.14\0(4)      & ---    & 3.96\0(6)& 3.907 &\07.16\0(4)       & ---    & \07.45(4)   & \07.399\\
\hline\hline
\end{tabular}
\end{center}
}
\end{table}
\end{turnpage}

\newpage
\begin{center}
\textbf{Figure 1}
\end{center}
\bigskip
Vapor pressure of carbon dioxide. {\large $\bullet$}~present results Grand Equilibrium method \cite{jadran}, {\small $\triangle$}~present results Gibbs ensemble method \cite{Gibbs}, {\small $\blacksquare$}~results by Zhang and Duan \cite{zhang2005}, {---}~reference EOS \cite{data}.
\bigskip
\begin{center}
\textbf{Figure 2}
\end{center}
\bigskip
Relative deviations of vapor-liquid equilibrium properties between simulation and reference EOS \cite{data} ($\delta z = (z_{\mathrm{sim}} - z_{\mathrm{eos}}) / z_{\mathrm{eos}}$). {\large $\bullet$}~present results Grand Equilibrium method \cite{jadran}, {\small $\triangle$}~present results Gibbs ensemble method \cite{Gibbs}, {\small $\blacksquare$}~results by Zhang and Duan \cite{zhang2005}, {---}~reference EOS \cite{data}.

\newpage

\newpage

\end{document}